\documentclass[twocolumn,multicolumn,aps,prb,showpacs]{revtex4}
\usepackage{graphicx}
\usepackage{dcolumn}
\usepackage{bm}
\usepackage{color}
\usepackage{latexsym}
\usepackage[T1]{fontenc}
\usepackage{amsmath}

\input{epsf}

\begin{document}

\preprint{APS/123-QED}

\date{\today}

\title{Shape- and orientation-dependence of surface barriers in single crystalline $d$-wave Bi$_2$Sr$_2$CaCu$_2$O$_{8+\delta}$}
\author{A.E. B\"{o}hmer, M. Konczykowski, C.J. van der Beek }
\affiliation{
Laboratoire des Solides Irradi\'{e}s, CNRS-UMR 7642 \& 
CEA/DSM/IRAMIS, Ecole Polytechnique, F 91128 Palaiseau cedex, France}

\begin{abstract}
Magneto-optical imaging and Hall-probe array magnetometry are used to measure the field of first flux 
entry, $H_p$,  into the same Bi$_2$Sr$_2$CaCu$_2$O$_{8+\delta}$ 
single crystal cut to different crystal thickness-to-width ratios $(d/w)$,
and for two angles $\alpha$ between the edges and the principal in-plane 
crystalline $(a,b)$ axes. At all temperatures, the variation with aspect 
ratio of $H_p$ 
is qualitatively well described by calculations for the so-called 
geometric barrier [E.H. Brandt, Phys. Rev. B {\bf 60}, 11939 (1999)]. 
However, the magnitude of $H_p$ is strongly enhanced due to the 
square shape of the crystal. In the intermediate temperature 
regime ($T \lesssim 50$ K) in which the Bean-Livingston barrier limits 
vortex entry, there is some evidence for a tiny crystal-orientation dependent enhancement 
when the sample edges are at an angle of 45$^{\circ}$ with respect to 
the crystalline axes, rather than parallel to them.
  
\end{abstract}
\maketitle

\section{Introduction}

Surface barriers are well-known to delay vortex penetration into high-$T_{c}$ 
cuprates (HTSC).\cite{Kopylov89,Kopylov90,Konczykowski91,Chikumoto92II,Koshelev91,Burlachkov93,Burlachkov94,Indenbom94,Indenbom94III,Zeldov94II,Zeldov95,Benkraouda96,Morozov97,Brandt2001,Brandt2001II,Gregory2001,Koshelev2001,Cabral2002,Wang2002,Conolly2008,Clem2008}
As a consequence, the field of first flux penetration $H_p$ may be significantly 
higher than the lower critical field $H_{c1} = 
\Phi_{0}/4\pi\mu_{0}\lambda^{2} \ln \kappa$ ($\lambda$ is the London
penetration depth, $\kappa \equiv \lambda/\xi$ with $\xi$ the coherence 
length, $\Phi_{0} = h/2e$ is the flux quantum, and $\mu_{0} = 
4\pi\times 10^{-7}$ Hm$^{-1}$). Among possible 
types of barriers, the Bean--Livingston (BL) barrier \cite{Koshelev91,Burlachkov93,Burlachkov94,deGennes62,Livingston64}
arises from the competition between the attraction of an entering vortex to the 
external surface (this can be described with an image vortex near the 
surface) and its repulsion by the screening (Meissner) current.  
The maximum value of the total free energy per unit 
length due to the introduction of a vortex, $\varepsilon_{0} \ln\left( H_{c}/ \sqrt{2} H_{a}\right)$, is 
attained when the latter is situated at a distance $\xi H_{c} / \sqrt{2} H_{a}$ from the 
superconductor-vacuum interface. Here,  $H_{a}$ is the applied magnetic 
field, $H_{c} = \sqrt{2} \kappa \varepsilon_{0}/\Phi_{0}$ is the thermodynamic 
critical field, and $\varepsilon_{0} = \Phi_{0}^{2}/4\pi\mu_{0}\lambda^{2}$ is the vortex 
line energy. Therefore, for fields close to the penetration field $H_{p} = \sqrt{2}H_{c}$, the BL barrier can be overcome 
by thermal activation. In long thin isotropic superconductors in 
parallel fields, this is thought to happen by the nucleation of a vortex loop in 
the sample bulk.\cite{Koshelev91,Burlachkov93} In layered superconductors such as Bi$_2$Sr$_2$CaCu$_2$O$_{8+\delta}$, 
the vortex loop may encompass a single \cite{Burlachkov94} or several 
pancake vortices.\cite{Koshelev91} Experiments supporting 
an important role of the BL barrier in HTSC  are the observation of 
asymmetric hysteresis loops accompanied by magnetic relaxation over the barrier, near the 
critical temperature $T_{c}$ in YBa$_{2}$Cu$_{3}$O$_{7-\delta}$,\cite{Konczykowski91} 
and at temperatures below 50 K in layered Tl$_2$Ba$_2$CaCu$_2$O$_{8+\delta}$ \cite{Kopylov89,Kopylov90} 
and Bi$_2$Sr$_2$CaCu$_2$O$_{8+\delta}$.\cite{Chikumoto92II,Zeldov95} 
Further evidence is the observation of 
the predicted sensitivity of the barrier to electron \cite{Chikumoto92II} and heavy-ion 
irradiation.\cite{Gregory2001,Koshelev2001} Namely, the presence of 
defects near the specimen surface perturb the screening current on the 
relevant length scale, larger than the coherence length $\xi$ but smaller 
than $\lambda$, thereby favoring vortex nucleation. 

Following theoretical predictions\cite{Iniotakis2008} of a possible crystal-orientation 
dependence of $H_{p}$ in HTSC, the BL barrier has recently returned to the center of attention.\cite{Leibowitz2009}
This orientational dependence would be due to the predominant 
$d_{x^{2}-y^{2}}$--symmetry of the superconducting gap in HTSC (for a review, 
see Ref.~\onlinecite{Tsuei2000}). In that case, quasi-particle 
scattering at surfaces oriented perpendicularly to the nodal 
directions give rise to  Andreev bound states 
(ABS).\cite{Hu94,Tanaka95,Kashiwaya2000} The quasi-particle 
current carried by these bound states is responsible for the 
modification of the field of first flux penetration.\cite{Iniotakis2008}  In 
Bi$_{2}$Sr$_{2}$CaCu$_{2}$O$_{8+\delta}$ the relevant surfaces are 
those that have a normal oriented near to $\alpha = 45^{\circ}$ with respect to the  principal $(a,b)$ crystalline 
axes.\cite{Kirtley96}

%
%
\begin{figure}[t]
\includegraphics[width=75mm]{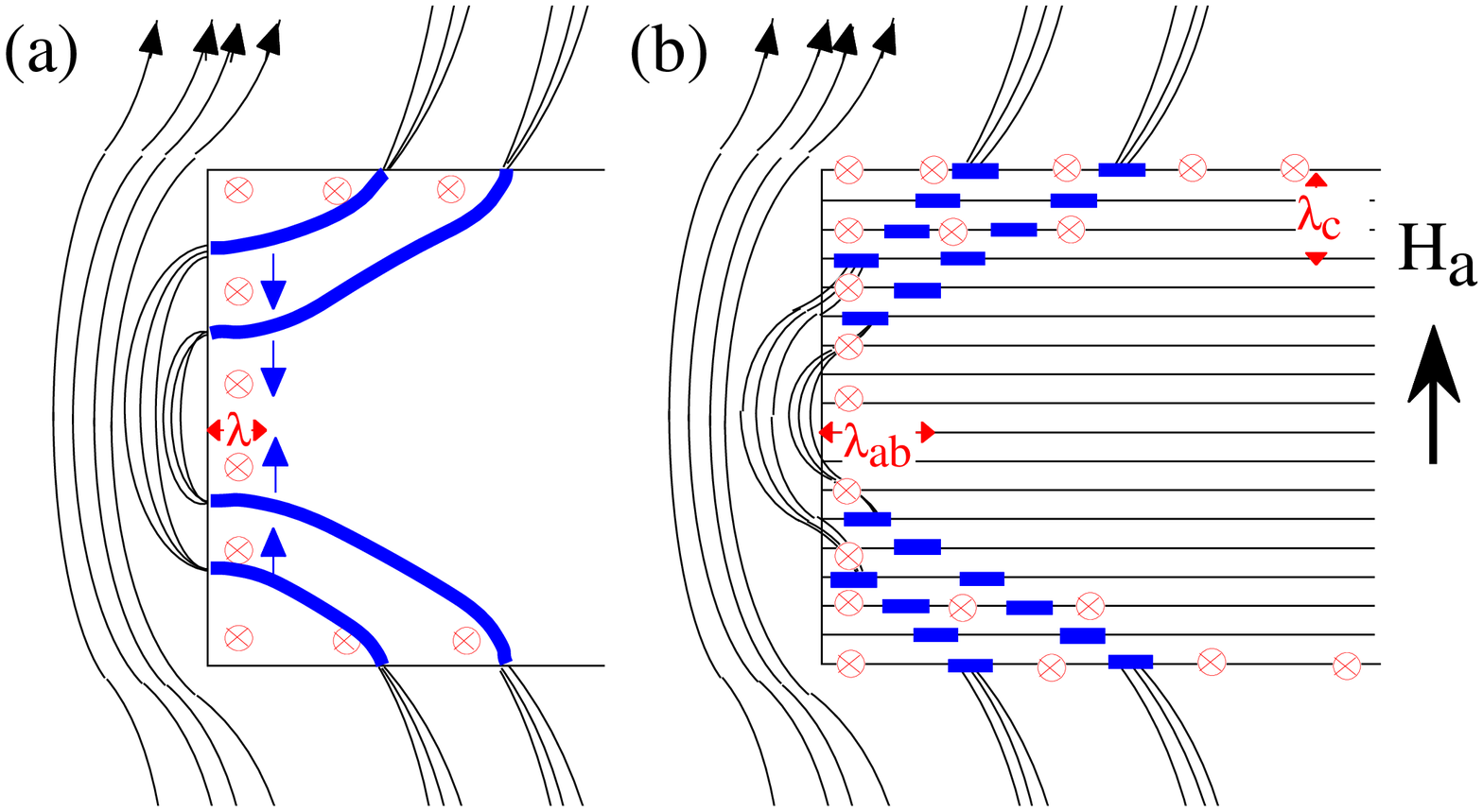}
\caption{(color online) Schematic diagram of vortex penetration in  (a) a ``continuous'' superconductor
in which layeredness is irrelevant and  (b)  in a layered 
superconductor; both materials are presumed to have a rectangular 
cross-section. Vortex lines (in blue) first penetrate at the corners, and then 
progress inwards to the sample equator. The direction of the 
screening current is indicated by the red circles. In 
case (a), there is no further barrier impeding the progression of the 
vortex lines (arrows); 
furthermore, the vortex currents run parallel to the lateral surface. Therefore, no  effect of the Bean-Livingston 
barrier superimposed on that of the geometrical barrier is to be 
expected. In (b), the progression of the vortex line is mediated by the 
periodic  entry of pancake vortices (blue rectangles); each of these 
should overcome the BL barrier.  The supercurrents associated with the pancakes run in a plane 
perpendicular to the lateral surface, therefore the geometrical barrier is enhanced by the 
Bean-Livingston barrier.}
\label{fig:Barriers-diagram}
\end{figure}

It should be noted that the BL barrier is effective only in 
the situation where the applied magnetic field is parallel to a large 
flat surface. In the opposite case of a 
superconducting platelet of rectangular cross-section in 
perpendicular magnetic field, flux entry is hindered by a barrier of geometrical origin.
\cite{Indenbom94,Indenbom94III,Zeldov94II,Zeldov95,Benkraouda96,Morozov97,Brandt2001,Brandt2001II,Cabral2002,Wang2002,Conolly2008}
The geometrical barrier comes from the competition between the line tension of a 
vortex cutting through the upper and lower ridges at a sample edge, and the Lorentz 
force due to the Meissner current.  The penetration field $H_{p}$ is 
reached when the total field at the sample equator attains $H_{c1}$; vortex sections 
penetrating from the top and the bottom ridge join, and the vortex line enters the bulk. 
A key experiment demonstrating the relevance of the geometrical 
barrier in Bi$_2$Sr$_2$CaCu$_2$O$_{8+\delta}$ is the measurement of 
the penetration field of crystals with similar aspect ratio $d/w$, 
but different shape: a prism-shaped\cite{Morozov97,Doyle97} or ellipsoidal\cite{Doyle98} crystal 
have a lower penetration field than a rectangular parallelepiped. 
Experiments reporting the accumulation of flux in the center of a 
superconductor sample immediately after first flux penetration cannot 
be taken as evidence for the presence of a geometrical barrier; such 
observations merely reflect the fact that the sample center corresponds to 
the location of minimal vortex free energy, a situation that stems from 
the macroscopic distribution of the Meissner current and is therefore true regardless 
of the origin of the barrier.

The flux density and current distributions accompanying vortex penetration into superconductors
of rectangular cross-section and varying thickness-to-width ratio $d/w$ was numerically calculated 
by Brandt.\cite{Brandt2001,Brandt2001II} The penetration field obtained from these calculations 
is well described by the formula 
	\begin{equation}
	H_p = H_{p}^{0,\infty} \tanh\left(\sqrt{\beta d/w}\right).
	\label{eq:tanh}
	\end{equation}
The constant $\beta$ was found to be $\sim 0.36$ for infinite strips 
(bars), and $\sim 0.67$ for rectangular cylinders; in the absence
of a BL barrier or vortex pinning, $H_{p}^{0,\infty} = 
H_{c1}$. Tentative agreement of Eq.~(\ref{eq:tanh}) with $H_{p}$ of 
micron-sized single crystalline Bi$_2$Sr$_2$CaCu$_2$O$_{8+\delta}$ 
was obtained for a few values of $d/w$ and temperature $T$ by Wang. \em et al. \rm  
,\cite{Wang2002} while Cabral \em et al. \rm  could fit their measured 
hysteresis loops to those calculated in 
Ref.~\onlinecite{Brandt2001II}, but only in the temperature interval 50 -- 60 K.\cite{Cabral2002}

The purpose of the present paper is twofold. In a first step, we shall 
verify the dependence of the field of first flux penetration in 
square-shaped Bi$_2$Sr$_2$CaCu$_2$O$_{8+\delta}$ single crystals 
on a wide range of aspect ratios $d/w$, over the entire temperature range from 20 K 
to $T_{c}$. This is achieved by cutting the same crystal into successively smaller squares.
These measurements establish the manner by which the 
Bean-Livingston barrier and magnetic relaxation influence $H_{p}$, and 
yield the geometrical factor $\beta$ for a crystal of square shape. 
The results are then used to verify whether any effect of 
crystalline orientation on Meissner screening and first flux 
penetration can be measured in single crystalline  
Bi$_2$Sr$_2$CaCu$_2$O$_{8+\delta}$. 

The expected orientation 
dependence \cite{Iniotakis2008} should actually be largest in layered 
superconductors such as Bi$_{2}$Sr$_{2}$CaCu$_{2}$O$_{8+\delta}$. In 
thin rectangular crystals of ``continuous'' superconductors such as 
YBa$_{2}$Cu$_{3}$O$_{7-\delta}$, there is no potential barrier 
stopping the progression of vortex lines into the interior once the 
geometrical barrier is overcome (Fig.~\ref{fig:Barriers-diagram}). 
Furthermore, the boundary condition imposes that the vortex current 
runs parallel to the crystal lateral surface.  In this situation, any 
effect of the ABS quasi-particle current should be confined to the 
volume near the crystal ridges.  In layered  superconductors, the 
progression of a vortex line is mediated by the successive 
penetration of pancake vortices. The entry of each pancake is 
counteracted by a potential barrier, since the pancake currents are 
always in a plane perpendicular to the lateral surface. These 
arguments hold for HTSC as long as the applied field is not exactly parallel to the $(a,b)$ plane, 
since the high crytalline anisotropy impedes vortex penetration along 
the $c$-axis. A possible contribution of ABS to the Bean-Livingston 
barrier in layered HTSC will come not only from the crystal 
ridges, but from all flat regions of the lateral 
surface that are oriented perpendicular to a nodal direction 
($\alpha =45^{\circ}$), and all wedge-shaped regions with bisector along the nodal 
direction.\cite{Iniotakis2007} In what follows, we find  a tiny enhancement 
of $H_{p}$ for $T < 50$ K, \em i.e. \rm, in the regime in which flux 
penetration is delayed by the Bean-Livingston barrier. This shows that, 
if any enhancement of the Bean-Livingston barrier due to ABS exists, it is
either much smaller than theoretically expected, or marred by physical damage 
at the crystal edges. At higher temperature, no enhancement is found.

\begin{figure}[t]
\includegraphics[width=0.35\textwidth]{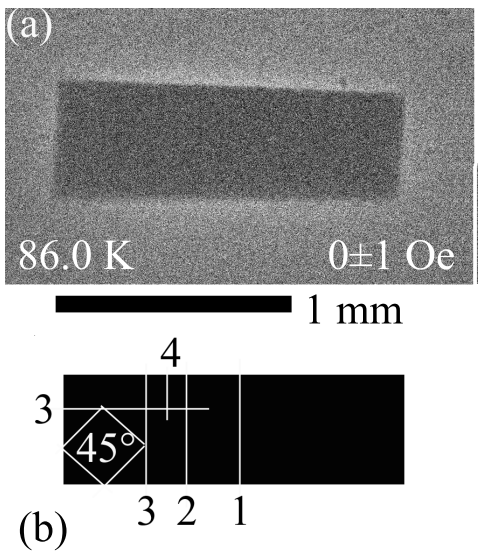}
\caption{(a) DMO image of Meissner screening by the initial 
Bi$_2$Sr$_2$CaCu$_2$O$_{8+\delta}$ single crystal, at 86.0 K 
and $H_{a} = 0$; the field modulation $\Delta 
H_{a} = 1$ Oe. (b)  Cutting scheme. The black rectangle 
represents the initial crystal, white numbered lines indicate successive cuts, in order.  }
\label{fig:MOimages-initial}
\end{figure}

\section{Experimental details}
\label{section:Experimental}

A rectangular Bi$_2$Sr$_2$CaCu$_2$O$_{8+\delta}$ 
single crystal is cut along its $a$ and $b$ axes from a seed rod grown 
by the travelling solvent floating zone method.\cite{MingLi2002} The 
crystal orientation was determined from the growth direction of the
rod,\cite{MingLi2002} and verified by the precession X-ray diffraction 
method.\cite{Buerger44}

\begin{figure}[t]
\includegraphics[width=0.45\textwidth]{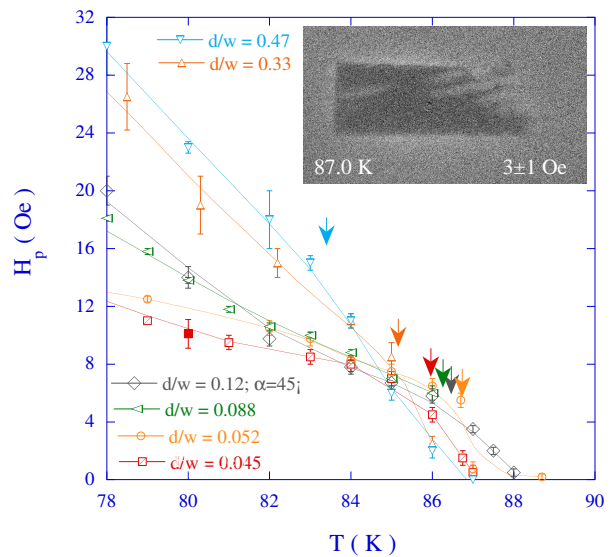}
\caption{(color online) Enlarged view of the penetration field data of 
Fig.~\protect\ref{fig:allfields} in the temperature regime close to 
$T_{c}$. Closed symbols denote data obtained by Hall probe 
magnetometry, open symbols are data measured using magneto-optical 
imaging. Data above 82 K are obtained using the differential 
magneto-optical method.  Arrows show the temperature above which flux penetration 
is influenced by crystal inhomogeneity. For example, the inset shows flux 
penetration in the initial crystal of 
Fig.~\protect\ref{fig:MOimages-initial} ($d/w = 0.045$) at 87.0 
K, measured by the DMO method at a field of 3 Oe, and a field 
modulation $\Delta H_{a} = 1$ Oe. The observed heterogeneity leads to 
the ``collapse'' of $H_{p}$ at $T = 86.8$ K. Lines are guides to the 
eye.}
\label{fig:Hp-zoom}
\end{figure}

\begin{figure}[b]
\includegraphics[width=75mm]{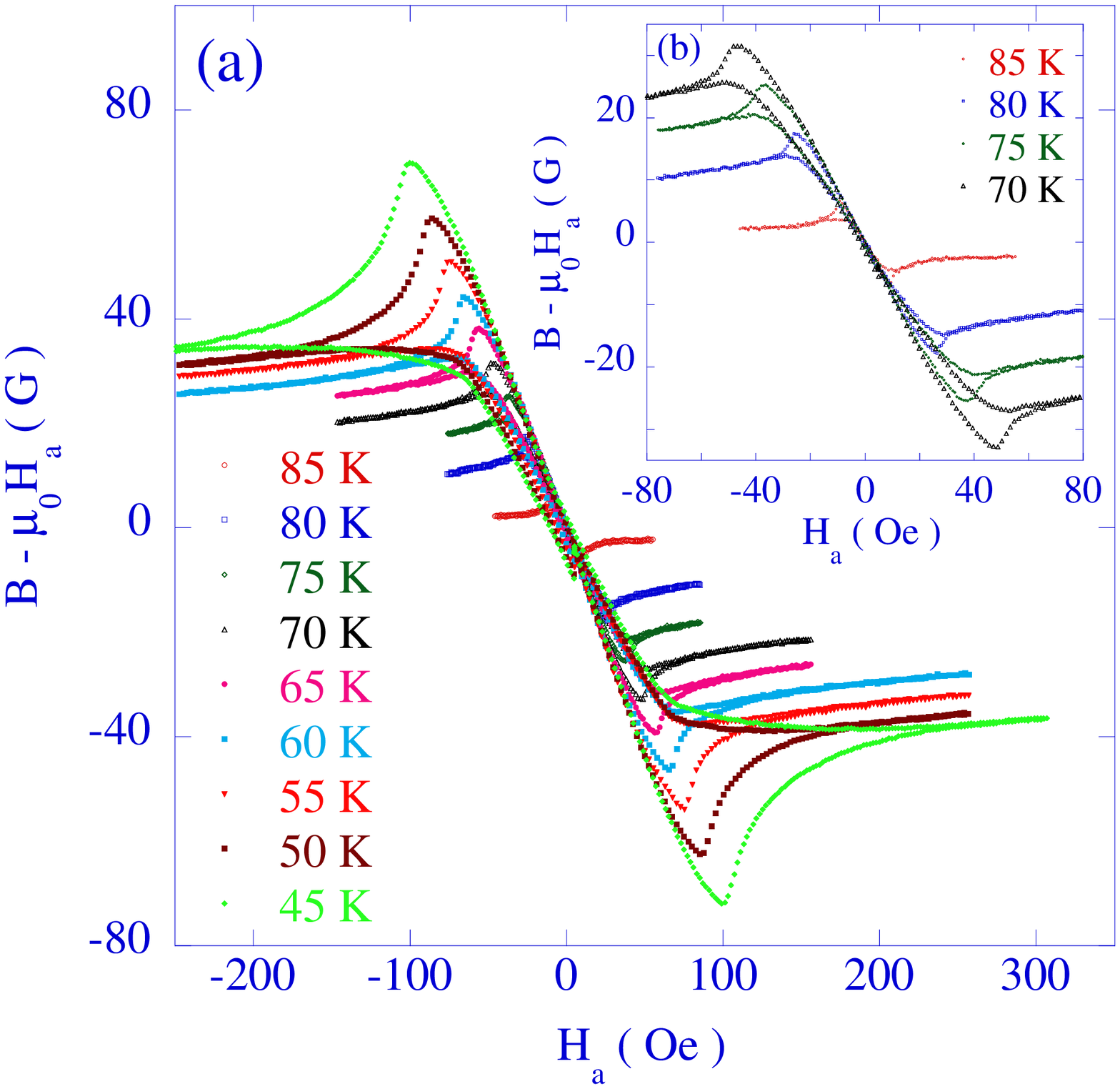}
\caption{(color online) Hysteresis loops of the magnetic flux density in the center 
of the crystal with $d/w = 0.052$, measured by a microscopic Hall 
probe array. The inset shows  an expanded view of the loops  
at high temperature. }
\label{fig:typicalhysteresis}
\end{figure}

Flux penetration below 80 K was imaged using direct magneto-optical imaging (MOI),
\cite{Dorosinskii92} while for higher temperatures the Differential 
Magneto-Optical (DMO) method\cite{Soibel99,Soibel2001} was used. In 
MOI, a ferrimagnetic garnet indicator with in-plane anisotropy is 
placed on top of the sample under study, and observed using a 
polarized light microscope. In all cases, the magnetic field, of magnitude $H_{a}$, is 
applied parallel to the smallest crystal dimension and the $c$-axis. The presence of a non-zero perpendicular 
component $B$ of the magnetic induction is revealed, by 
virtue of the Faraday effect of the garnet, as a non-zero intensity of 
reflected light when the polarizers of the microscope are (nearly) 
crossed (see Fig.~\ref{fig:homogeneous}). Images are acquired by an 
automated procedure, which also ramps the magnetic field, at the same 
rate, $\mu_{0}dH_{a}/dt = 5 \times 10^{-4}$ 
Ts$^{-1}$, for all experiments. In DMO, an image is taken 
at an applied field $H_{a}$; the field is then decreased to $H_{a} - 
\Delta H_{a}$ (with $\Delta H_{a} = 1$ Oe), and another image is 
acquired. The latter image is then subtracted from the former. The 
procedure is repeated 100 times, whence the 100 differential images 
are averaged 
(see Fig.~\ref{fig:MOimages-initial}a, and the inset to 
Fig.~\ref{fig:Hp-zoom}). The use of the DMO technique was limited to 
temperatures above 82 K, at which the field modulation $\Delta H_{a}$ 
is not screened once the magnetic field exceeds $H_{p}$. 
The field of first flux penetration was determined 
visually from both the direct and the differential magneto-optical 
images, taken at successive values of $H_{a}$, from 
the appearance of a non-zero $B$ within the sample contours (Fig.~\ref{fig:homogeneous}). 
A more precise value could be obtained by calibrating the local intensity of the 
light; this yields a map of the local induction from which curves of 
the local magnetic hysteresis were deduced. Similar curves were obtained using a micron-sized Hall probe 
array.\cite{Zeldov94II} Examples of these are shown in 
Fig.~\ref{fig:typicalhysteresis}. The penetration field was then 
estimated from the peak in the hysteresis curves ($\mu_{0}dH_{a}/dt = 1 
\times 10^{-4}$ Ts$^{-1}$ for these measurement).

\begin{figure}[b]
\includegraphics[width=0.5\textwidth]{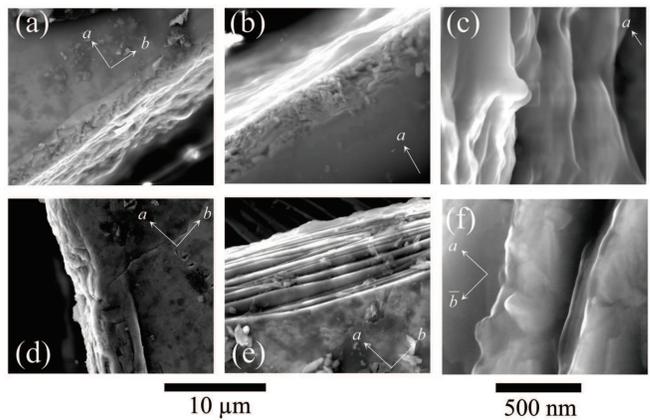}
\caption{Scanning electron micrographs of crystal edges cut at 
$90^{\circ}$ (a-c) and at $45^{\circ}$ with respect to the principal [100] and [010] crystal 
axes (d-f).  Images (a,b;d,e) have the same magnification, as do images 
(c,f).  }
\label{fig:Edges}
\end{figure}

In order to avoid any artefacts due to tiny variations in sample 
composition\cite{Li94} or doping, the same 
Bi$_2$Sr$_2$CaCu$_2$O$_{8+\delta}$  crystal is chosen for all further 
experiments. The selected  crystal, with $T_{c} = 88.7 \pm 0.5
$K, had initial dimensions  $(l = 1400) \times  (w = 
500) \times  (d = 28)$ $\mu$m$^3$, and was checked  using the differential 
magneto-optical (DMO) method\cite{Soibel99,Soibel2001} to be as homogeneous 
as possible: macroscopic and mesoscopic 
inhomogeneities are known to affect $H_{p}$.\cite{Conolly2008,Avraham2009} 
Nevertheless, some inhomogeneities arising from the crystal growth 
method remained.\cite{Yasugaki2003} Fluctuations in crystal growth rate 
result in small spatial variations of the composition, with a 
arc-shaped topology reminiscent of the floating-zone meniscus (inset of 
Fig.~\ref{fig:Hp-zoom}). The associated spatial variation of the local $T_{c}$ 
leads to preferential flux penetration in the arc-shaped regions, 
above the temperatures marked by arrows in Fig.~\ref{fig:Hp-zoom}.

\begin{figure}[t]
\includegraphics[width=0.48\textwidth]{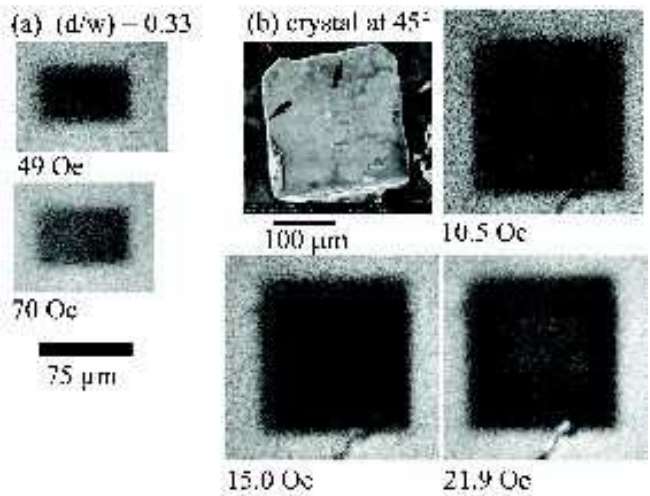}
\caption{(a) Flux penetration into the crystal with $d/w = 0.33$, delimited by lines ``3'', ``3'', 
and ``2'' in Fig.~\protect\ref{fig:MOimages-initial}(b), at 70.0 K, and two fields below ($\mu_{0}H_{a} = 4.9$ mT) and 
above ($\mu_{0}H_{a} = 7.0$ mT) $H_{p}$, respectively. (b) Scanning 
electron micrograph of the crystal cut at $45^{\circ}$, with arrows 
showing the areas enlarged in Fig.~\protect\ref{fig:Edges} (d,e); the 
subsequent three images show flux 
penetration into this crystal as visualized by MOI at 80.0 K and 
$H_{a} = 10.5$ Oe $< H_{p}$, 15.0 Oe $\approx H_{p}$, as well as $H_{a}= 21.9$ Oe$ > H_{p}$.}
\label{fig:homogeneous}
\end{figure}

For the measurements of $H_{p}$ as function of $d/w$, the same roughly square shape 
is retained to eliminate the influence of sample 
shape, or ambiguities related to the difference between length and 
width. Using a W wire saw with 1 $\mu$m SiC abrasive grains, the initial crystal, with 
$d/w = 0.045$, is progressively cut into five ever smaller squares, down 
to $d/w = 0.47$, according to the scheme of Fig.~\ref{fig:MOimages-initial}b.
In the final step,  one of the intermediate squares is cut at 45$^{\circ}$ with 
respect to the crystal axes; it had an aspect ratio of $d/w=0.12$. 
Fig.~\ref{fig:Edges} shows scanning electron micrographs of the edges 
obtained following the cuts. Typically, different layers of the crystal split at 
slightly different distances $x$ from the crystal center, yielding an edge 
roughness $(\langle x^{2} \rangle - \langle x\rangle^{2} )^{1/2} \sim 2-3$ 
$\mu$m. The edges of individual layers are smooth on the scale of the 
penetration depth, $\sim 200$ nm, with occasional protrusions. The 
local orientation of the edge meanders about the global orientation of 
the crystal, with a rms deviation of 10--25$^{\circ}$, depending on 
the investigated edge. No systematic morphological differences could be 
observed between [100] or [010] ``anti-nodal'' and [110] ``nodal'' interfaces. 
The crystal edges therefore seem to be suitably described by the 
model advocated in Ref.~\onlinecite{Iniotakis2007}.

\begin{figure}[b]
\includegraphics[width=0.45\textwidth]{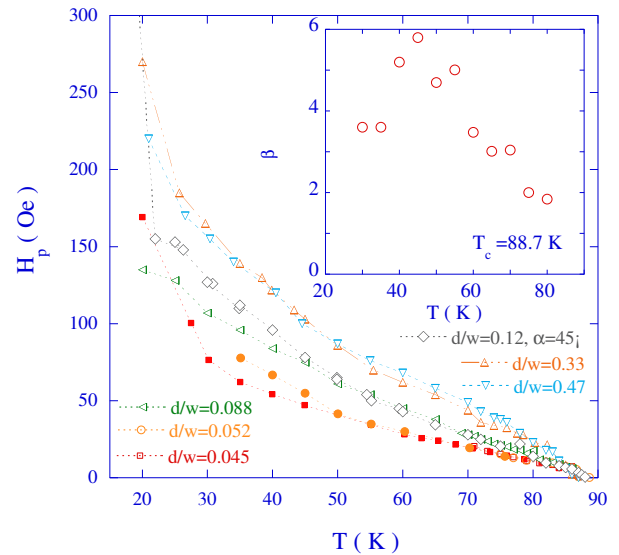}
\caption{(color online) The penetration field as evaluated from MO-imaging 
(open symbols) and Hall probe measurements (full symbols). The aspect ratio is 
$d/w=0.47,\,0.33,\,0.12 (45^{\circ}),\,0.088,\,0.052,\,0.045,$ from the upper to the lower 
curves. The inset shows the experimentally determined factor $\beta$ 
as function of temperature.}
\label{fig:allfields}
\end{figure}

\section{Results} 

The main panel of Fig.~\ref{fig:Hp-zoom} shows $H_{p}$ of the different crystals cut from the initial sample in the vicinity of 
$T_{c}$. One systematically observes a linear decrease of $H_{p}(T)$ 
upon approaching $T_{c}$, followed by a steep descent to $H_{p}=0$ 
(see also Ref.~\onlinecite{Brawner}). DMO imaging shows that this drop is the consequence of flux 
penetration into areas of lower $T_{c}$, and is therefore the 
hallmark of sample inhomogeneity. A  similar depression of the field of first flux 
penetration was obtained by introducing heavy ion-irradiated regions.\cite{Avraham2009}  
In what follows, we shall discard data taken in the high temperature regime of 
inhomogeneous flux penetration.

\begin{figure}[t]
\includegraphics[width=0.45\textwidth]{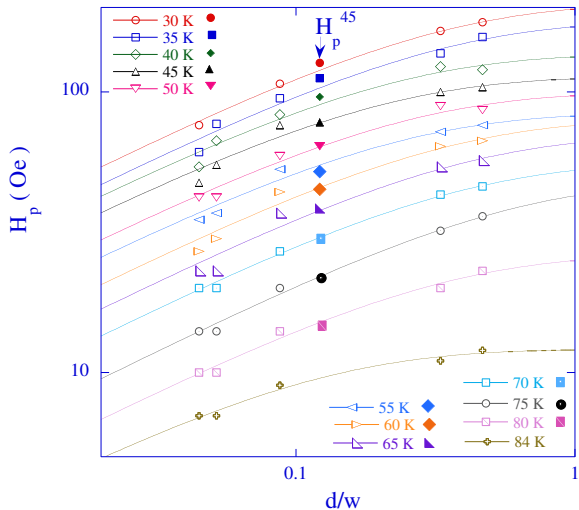}
\caption{(color online) Double-logarithmic plot of the penetration field of single crystalline 
Bi$_{2}$Sr$_{2}$CaCu$_{2}$O$_{8}$  as a function of aspect ratio for 
different temperatures. Open symbols show $H_{p}^{0}$ for the case 
where the edges are parallel to the principal axes. Drawn lines are fits to 
Eq.~(\protect\ref{eq:tanh}). Full symbols indicate $H_{p}^{45}$ 
(edges at $\alpha = 45^{\circ}$ with respect to the principal crystal 
axes).}
\label{fig:tanh}
\end{figure}

Fig.~\ref{fig:allfields} collects the penetration fields for all 
temperatures and all thickness/width ratios. As expected from the demagnetizing effect, 
the field of first flux entry strongly depends  on the aspect ratio of the sample. 
Wide thin platelets have a lower penetration field than narrow thick ones. This shows that the 
local induction near the edge is mainly determined by the shape of the 
crystals, and that edge roughness and protrusions 
(Fig.~\ref{fig:Edges}) do not play a 
determining role in defining $H_{p}$. The measured dependence of $H_{p}$ on $d/w$, 
presented in Fig.~\ref{fig:tanh}, is very well described by 
Eq.~(\ref{eq:tanh}). However, the experimental values $\beta$, ranging 
between $\sim 2 - 6$ (inset to Fig.~\ref{fig:allfields}), are significantly 
larger  than expected. The prefactor $H_{p}^{0,\infty}$ determined from the fits is 
presented in Fig.~\ref{fig:Hp-infinite}. Above $T \sim 
45$ K, it corresponds to $H_{c1}(T)$ of the $d$-wave superconductor \cite{Radtke96} in the limit $d/w \rightarrow 
\infty$, with the correct\cite{MingLi2002}  zero-temperature extrapolated 
$H_{c1}^{0}(0) = 180$ Oe and $T_{c} = 88.7$ K as obtained from 
DMO. Previously attempted comparisons to the 
two-fluid relation,\cite{Cabral2002} $H_{p}^{0,\infty}(T) = H_{p}^{0,\infty}(0)\left[ 
1-\left(T/T_c\right)^4\right]$ yield an anomalously low value $H_{c1}(0) \approx 
100$ Oe. 

We note the observation of thermally activated flux entry at 
all temperatures, indicative of a role of the BL barrier. 
In particular, the width of the hysteresis loops of 
Fig.~\ref{fig:typicalhysteresis} progressively narrows 
as $T$ increases, until the loops are all but closed for $T \lesssim 
T_{c}$. As in Ref.~\onlinecite{Gregory2001}, the temperature 
dependence of the penetration field in the regime 40 -- 50 K is in 
qualitative agreement with the theory of 
Ref.~\onlinecite{Burlachkov94}, which has 
\begin{equation}
    H_{p} = \frac{H_{c}}{\sqrt{2}} \left( \frac{t_{0}+t}{\tau} \right)^{-k_{B}T/\varepsilon_{0}s}.
    \label{eq:Hp-with-relaxation}
    \end{equation}
The relaxation rate is observed to be $S \sim -0.03$.  However, if one 
attempts a quantitative fit by inserting this value, which 
should correspond to $S = dH_{p}/d\ln t \equiv 
k_{B}T/\varepsilon_{0}(T)s$, into Eq.~(\ref{eq:Hp-with-relaxation}), 
one obtains $H_{p}$--values that, above 50 K, 
fall vastly short of the measured ones. Therefore, vortex entry 
above this temperature is not limited by the BL barrier (see also 
Ref.~\onlinecite{Zeldov95}). Below $T \sim 30$ K, the shape of the hysteresis loops 
and the observation of remnant flux clearly indicate the effect of flux 
pinning. $H_{p}$ shows an associated steep exponential 
increase as $T$ is lowered.

The fits to Eq.~(\ref{eq:tanh})  yield a set of 
master curves allowing us to assess the dependence 
of $H_{p}$ on the relative orientation of the sample edges with respect to the 
principal crystal axes. For this, we present, in Fig. \ref{fig:Hp-infinite}, the $H_{p}^{45}$ data measured 
on the sample with $d/w = 0.12$, cut at  $45^{\circ}$.
These are compared to the interpolated values of $H_{p}^{0}$ for a hypothetical sample with the same 
aspect ratio, cut parallel to the principal axes ($\alpha = 0$). At 
temperatures above 50 K  where the Bean-Livingston barrier is 
inoperative,\cite{Zeldov95} we find $H_{p}^{45}$ to be slightly 
smaller than $H_{p}^{0}$. However, in the regime where the 
Bean-Livingston barrier impedes pancake vortex entry into the 
crystal, there is a small enhancement $H_{p}^{45}/H_{p}^{0}$ of up to about 
1.25. This enhancement increases with decreasing temperature.

\section{Discussion}

We first discuss the enhancement of the penetration field by the BL 
 barrier and the effect of sample shape for $\alpha = 0$. The 
dependence of $H_{p}$ on $d/w$ is well described by 
Eq.~(\ref{eq:tanh}) at all temperatures, not only at those where the barrier is of 
purely geometric origin. This is a consequence of the macroscopic 
distribution of the Meissner current, which depends only on sample 
shape and the value of the penetration depth. At temperatures below 
50 K, pancake vortex entry is hindered additionally by the BL 
barrier, which leads to a larger penetration field.\cite{Zeldov95} This 
enhancement is manifest through the increase of 
$\beta$ as temperature is lowered. As temperature increases, pancake 
vortex penetration is accelerated due to thermal activation over the 
BL barrier. However, the field of first flux penetration 
can never drop below that of the geometrical barrier, since below this 
field $H < H_{c1}$ at the crystal equator and the existence of 
vortices in the superconductor is thermodynamically unfavorable. 
Once $H_{p}$ is overcome, thermal activation of pancakes leads to the 
rapid increase of the local induction at the sample boundary to 
values above $H_{c1}$, and the concommitant closing of the 
magnetic hysteresis loops.

Even in the high temperature regime where the BL
barrier has no effect whatsoever, the experimentally found  $\beta$ is 
much larger than that found in the calculations of 
Refs.~\onlinecite{Brandt2001} and \onlinecite{Brandt2001II}.  Such a large 
value of $\beta$ expresses the fact that in thinner crystals, the 
measured penetration field is larger than what would be expected from 
calculations including the effect of the geometrical barrier only.
An explanation is that, apart from opposing pancake entry from the side surfaces, 
the Bean-Livingston barrier opposes initial flux penetration across the top and 
bottom sample ridges, 
a finite size-effect that is expected to be 
more important for thinner crystals.
Furthermore, $\beta$ is enhanced because of the square shape of the sample 
surface perpendicular to the field direction. Although this is manifestly 
closer to the disk than to the strip geometry, the corners of a square superconductor 
are responsible for enhanced screening with respect to the disk, yielding a larger 
penetration field. The effect is similar to that of the  
ridges of a platelet of rectangular cross-section, which are 
responsible for additional screening when compared to an ellipsoidal superconductor of the 
same $d/w$. By extrapolating to $T_{c}$, at which all effects of the 
Bean-Livingston barrier have vanished, we find, experimentally, 
that $\beta \sim 1$ for a square superconductor with rectangular cross-section.
Note that taking an incorrect value of $\beta$ may result in large error when 
determining of the first critical field $H_{c1}$ from $H_{p}$ of thin 
superconducting platelets.

\begin{figure}[t]
\includegraphics[width=0.45\textwidth]{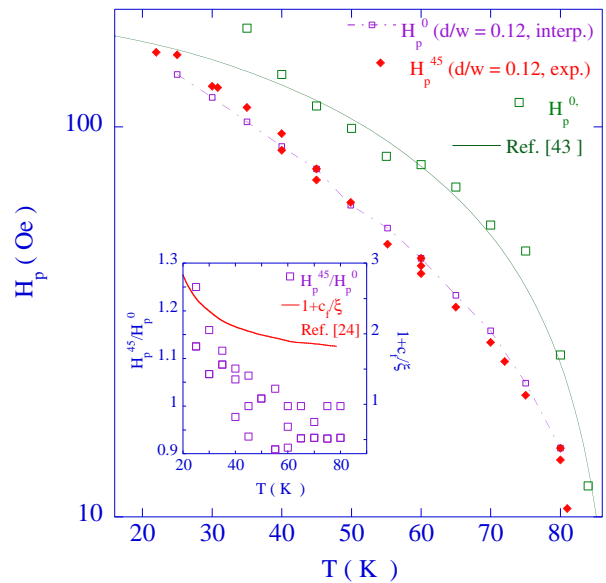}

\caption{(color online) Plot of the penetration field $H_{p}^{45}$ of a 
Bi$_{2}$Sr$_{2}$CaCu$_{2}$O$_{8}$ crystal with the edges 
cut at $\alpha = 45^{\circ}$ with respect to the main crystal axes 
(\color{red} $\diamondsuit$ \color{black}), compared to the interpolated 
values $H_{p}^{0}$ for a crystal with the same $d/w = 0.12$, but $\alpha = 0$ 
(\color{magenta} \fbox{\hspace{0.3mm}}{} \color{black}).
Also shown is $H_{p}^{0\infty}(T)$ for a crystal with $d/w 
\rightarrow \infty$, $\alpha = 0$, extracted from the fits in 
Fig.~\protect\ref{fig:tanh}
(\color{green} \fbox{\hspace{0.mm}}{} \color{black}); the drawn line 
is a fit to the theoretical temperature dependence of $H_{c1}$ 
for a $d$-wave superconductor,\protect\cite{Radtke96}
with $H_{c1}(0)= 180$ Oe and $T_{c} = 88.7$ K. The inset shows the 
experimental (lefthand axis) and the theoretically expected 
enhancement (righthand axis) of the field of first flux penetration for $\alpha = 
45^{\circ}$ with respect to $\alpha = 0$. }
\label{fig:Hp-infinite}
\end{figure}

We now turn to the relative enhancement $H_{p}^{45}/H_{p}^{0}$ of the 
field of first flux penetration for $\alpha = 45^{\circ}$, \em i.e. \rm where the crystal edges 
are perpendicular to the nodal directions of the superconducting order 
parameter in Bi$_{2}$Sr$_{2}$CaCu$_{2}$O$_{8+\delta}$. The 
observation of an enhancement in the sole temperature regime where the 
BL barrier is relevant is in qualitative agreement with the theory of 
Iniotakis, Dahm, and Schopohl. The theoretically predicted 
enhancement $H_{p}^{45}/H_{p}^{0} \sim  1 + c_{f}/\xi$ increases as temperature 
decreases (see Inset to Fig.~\ref{fig:Hp-infinite}). Its physical 
origin is that, in superconductors where the order parameter $\Delta$ undergoes a sign change 
on any given Fermi surface sheet, interference between quasiparticles 
with a relative phase difference of $\pi$ will result in the occurence 
of Andreev bound states (ABS) on those parts of the superconductor-vacuum 
interface that are perpendicular, or nearly perpendicular, to the $k$-space 
direction for which $\Delta$ vanishes. For 
Bi$_{2}$Sr$_{2}$CaCu$_{2}$O$_{8+\delta}$ with $d_{x^{2}-y^{2}}$-wave symmetry of the 
order parameter,\cite{Tsuei2000} these ``nodal'' directions correspond 
to $(\pm \pi, \pm \pi$), \em i.e.\rm ~at $\alpha = 45^{\circ}$ with 
respect to the principal 
crystalline axes.\cite{Tsuei2000}  The ABS have a spatial extent of several times $\xi$ 
in the direction perpendicular to the interface, 
and are accompanied by a local reduction of $|\Delta|$. 
Translational invariance along the interface allows these 
states to carry current. When a magnetic field is applied to the 
superconductor, this ``anomalous'' paramagnetic current adds to the diamagnetic 
screening current, and to the ordinary paramagnetic current resulting 
from the Doppler shift of quasiparticle states.\cite{Tinkham} In the 
region near a nodal surface, the total Meissner current 
is therefore reduced with respect to surfaces that are perpendicular 
to the $k$-space direction where the 
gap is maximum;\cite{Fogelstrom97} hence, the magnetic field 
should penetrate somewhat further into the superconductor.\cite{Iniotakis2008} 
The presence of a vortex near the nodal interface should also affect the Andreev bound 
states,\cite{Graser2004} with the result
of somewhat reducing the anomalous current.\cite{Iniotakis2008} The BL 
barrier on the nodal surface must therefore be evaluated by the 
addition of the standard free energy contributions from the vortex-Meissner 
current repulsion and the attraction of the vortex to the surface; and 
the ``anomalous'' contributions arising from the extra repulsion due 
to the augmented field near the surface, and from the attraction 
between the vortex current and the bound state 
current.\cite{Iniotakis2008} The net result is an enhancement of the 
barrier maximum, and a shift of its location towards the bulk of the 
superconductor. The enhancement of the BL barrier should be 
maximum when the orientation of the superconductor-vacuum interface 
is perpendicular to the nodal direction (and nil when the interface 
is parallel to the nodal direction), and be temperature dependent. For example, at $T/T_{c} = 0.2$, an 
increase of $H_{p}$ by a factor of 4 to 5 is expected on surfaces 
oriented at $\alpha = 45^{\circ}$, with respect to the situation where $\alpha = 0$.\cite{Iniotakis2008}

The observation of an enhancement of $H_{p}$ for the crystal with the edges cut perpendicular to the 
nodal directions seems to confirm the prediction by Iniotakis \em et al. \rm of an 
orientation-dependent $H_{p}$ due to a more important effect of Andreev 
bound states in this case.\cite{Iniotakis2007,Iniotakis2008}  The inset to
Fig.~\ref{fig:Hp-infinite} shows that the magnitude of the 
experimentally measured effect is, however, far less than predicted.\cite{Iniotakis2008}
The prime factor that may influence the experimental result is the roughness of the edges cut by the wire 
saw. The undulating character of the crystal edges will entail a reduction of the local density of bound states, 
and should strongly diminish the total anomalous (paramagnetic) Meissner 
current. As for the observed wedge-shaped protrusions, recent work has 
shown even if the occurence of ABS depends sensitively on 
quasiparticle reflection at the boundary as well as the shape and 
precise orientation of the latter,\cite{Iniotakis2005} they are still expected 
as long as the wedge bisector is aligned close to a nodal direction.\cite{Iniotakis2007} 
However, the localised quasiparticles near to such protrusions will not contribute to the total Meissner 
current, and will have no effect on the Bean-Livingston barrier. 
In summary, the  area on which the Bean-Livingston barrier is 
effectively enhanced is much smaller than the total area of the 
lateral edges. Only a fraction of pancakes composing an entering 
vortex line will effectively be subjected to the influence of local 
ABS, entailing the decrease of $H_{p}$ 

Finally, even in those parts of the surface where ABS would 
be fully developped, pancake vortices enter the crystal by thermal 
activation over the barrier. The implies that pancakes are nucleated 
at a distance that is farther removed from the edge than the 
predicted $\xi H_{c} / \sqrt{2} H_{a}$.\cite{Burlachkov94} 
In Ref.~\onlinecite{Iniotakis2008}, the magnitude and temperature dependence of 
the enhancement factor $1 +c_{f}/\xi$ strongly depends on the distance 
of the vortex from the crystal edge. Any enhancement of this distance 
due to thermal activation would imply a much smaller modification of the Meissner 
current, and a significant reduction of the enhancement factor.

\section{Conclusions}

Summarizing, we have measured the field of first flux entry, $H_{p}$, for  
samples of different aspect ratio and orientation, cut from the same optimally doped 
Bi$_{2}$Sr$_{2}$CaCu$_{2}$O$_{8}$ single crystal. We find that at 
all temperatures, the penetration field of the  
square--shaped crystals is satisfactorily described by 
Eq.~(\ref{eq:tanh}). The numerical factor  $\beta$ 
is determined jointly by the shape of the sample surface perpendicular 
to the field direction and by the effect of the Bean-Livingston barrier, which, in layered 
superconductors, counteracts the penetration of individual pancake 
vortices. For the case of a square superconductor of rectangular 
cross-section, without the influence of a Bean-Livingston barrier, 
it is experimentally found that $\beta \sim 1$.

Moreover, we have investigated the enhancement of $H_{p}$ when the edges are 
perpendicular to the nodal directions of the order parameter, \em i.e. \rm at 
$45^{\circ}$ with respect to the crystalline $a$ and $b$ axes. If 
discernible, the measured enhancement is tiny and occurs only in the temperature 
interval in which the Bean-Livingston barrier increases the field of 
first flux penetration over what would be expected from sole 
geometrical effects. The results are discussed in the framework of 
possible Andreev bound states 
present at the nodal surfaces of the $d$-wave  
superconductor.\cite{Iniotakis2008} Notably, the small value of the 
enhancement in comparison to theoretical prediction is ascribed to 
edge roughness and thermal activation of pancake vortices over the 
Bean-Livingston barrier. 

\section*{Acknowledgements} We wish to acknowledge O. Mula, H. Martin, 
M. Goupil, M. Girardin, and J. Arancibia who participated in 
part of the work, and B. Mansart and P.A. Albouy for help with the X-ray 
diffraction measurement, and D. Caldemaison for the SEM micrographs. 
We are also indebted to E.H. Brandt, V.B. Geshkenbein, T. Shibauchi, T. Tamegai, U. Welp, 
and E. Zeldov for discussions.

\end{document}